\RequirePackage{rotating}
\documentclass[manuscript, screen]{acmart}
\usepackage{amssymb}
\usepackage{multicol}
\usepackage{multirow}
\usepackage{rotating}
\usepackage[inline]{enumitem}

\usepackage{booktabs}
\usepackage{graphics}
\usepackage{subcaption}
\makeatletter \def\mdseries@tt{m} \makeatother
\usepackage[newfloat,draft=true]{minted}

\usepackage{todonotes}
\newcommand{\am}{\todo[author=AM, color=red!20!white,inline, disable=True]}
\newcommand{\ms}{\todo[author=MS, color=blue!20!white,inline, disable=True]}

\usepackage{comment}
\usepackage{color}

\begin{document}

\setcopyright{acmcopyright}
\acmJournal{TSC}
\acmYear{2019} \acmVolume{1} \acmNumber{1} \acmArticle{1} \acmMonth{1} \acmPrice{15.00} \acmDOI{10.1145/3352572}

\title{A Practical Guide for the Effective Evaluation of Twitter User Geolocation}

\author{Ahmed Mourad}
\orcid{0000-0002-9423-9404}
\affiliation{%
  \institution{School of Computer Science and Information Technology, RMIT University}
  \streetaddress{124 La Trobe Street}
  \city{Melbourne}
  \state{VIC}
  \postcode{3000}
  \country{Australia}}
\email{ahmed.mourad@rmit.edu.au}

\author{Falk Scholer}
\orcid{0000-0001-9094-0810}
\affiliation{%
  \institution{School of Computer Science and Information Technology, RMIT University}
  \streetaddress{124 La Trobe Street}
  \city{Melbourne}
  \state{VIC}
  \postcode{3000}
  \country{Australia}}
\email{falk.scholer@rmit.edu.au}

\author{Walid Magdy}
\orcid{0000-0001-9676-1338}
\affiliation{%
  \institution{School of Informatics, The University of Edinburgh}
  \streetaddress{10 Crichton Street}
  \city{}
  \state{Edinburgh}
  \postcode{EH8 9AB}
  \country{United Kingdom}}
\email{wmagdy@inf.ed.ac.uk}

\author{Mark Sanderson}
\orcid{0000-0003-0487-9609}
\affiliation{%
  \institution{School of Computer Science and Information Technology, RMIT University}
  \streetaddress{124 La Trobe Street}
  \city{Melbourne}
  \state{VIC}
  \postcode{3000}
  \country{Australia}}
\email{mark.sanderson@rmit.edu.au}

\renewcommand\shortauthors{Mourad, A. et al}

\begin{abstract}
Geolocating Twitter users---the task of identifying their home locations---serves a wide range of community and business applications such as managing natural crises, journalism, and public health. Many approaches have been proposed for automatically geolocating users based on their tweets; at the same time, various evaluation metrics have been proposed to measure the effectiveness of these approaches, making it challenging to understand which of these metrics is the most suitable for this task. In this paper, we propose a guide for a standardized evaluation of Twitter user geolocation by analyzing fifteen models and two baselines in a controlled experimental setting. Models are evaluated using ten metrics over four geographic granularities. We use rank correlations to assess the effectiveness of these metrics.

Our results demonstrate that the choice of effectiveness metric can have a substantial impact on the conclusions drawn from a geolocation system experiment, potentially leading experimenters to contradictory results about relative effectiveness. We show that for general evaluations, a range of performance metrics should be reported, to ensure that a complete picture of system effectiveness is conveyed. Given the global geographic coverage of this task, we specifically recommend evaluation at micro versus macro levels to measure the impact of the bias in distribution over locations. Although a lot of complex geolocation algorithms have been applied in recent years, a majority class baseline is still competitive at coarse geographic granularity. We propose a suite of statistical analysis tests, based on the employed metric, to ensure that the results are not coincidental~\footnote{The code for the evaluation framework detailed in this article can be found on: https://bitbucket.org/amourad/geoloceval.git}.
\end{abstract}

%
%

\ccsdesc[500]{General and reference~Evaluation}

%
%

\keywords{Twitter, User Geolocation, Effective Evaluation, Statistical Analysis}

\maketitle

\section{Introduction}
\label{sec:intro}

Geolocating Twitter users is needed in many social media-based applications, such as identifying geographic lexical variation~\cite{eisenstein2010latent,han2014text}, managing natural crises~\cite{kryvasheyeu2015performance}, gathering news~\cite{zubiaga2013curating,schwartz2015editorial,liu2016reuters}, and tracking epidemics~\cite{dredze2013carmen,broniatowski2013national}. While users can record their location on their profile, \citet{hecht2011tweets} reported that more than $34\%$ record fake or sarcastic locations. Twitter allows users to GPS locate their content, however, \citet{han2014text} reported that less than $1\%$ of tweets are geotagged. Inferring user location is therefore an important field of investigation.

Each geolocation application has different needs, which might require evaluation from several perspectives. However, current evaluation practices focus on a few measures introduced by \citet{eisenstein2010latent}. These measures were shown to be biased towards densely populated (urban) locations~\cite{mourad2017language}, e.g. the accuracy over urban locations will dominate the overall measure. Such measures may be unsuitable to evaluate applications that treat urban and rural locations with the same degree of importance: e.g. searching for sources to cover local news~\cite{starbird2012learning,schwartz2015editorial,liu2016reuters}, monitoring natural disasters in rural areas~\cite{kryvasheyeu2015performance}, or tracking epidemics in rural cities~\cite{dredze2013carmen}.

Moreover, evaluation at multiple levels of geographic granularity is not widely used despite it being required by some applications. For instance, \citet{diakopoulos2012finding}, in determining requirements from journalists for identifying eyewitnesses from social media, found that aggregating predicted eyewitness location at different scales was requested, e.g. city, state or country. Similarly, \citet{dredze2013carmen} presented a geolocation prediction system (Carmen) for influenza surveillance, which predicts a structured location at different granularities.

The evaluation of geo-inference methods is affected by many factors, such as dataset availability, pre-processing, ground-truth construction, geographic coverage, and how the earth is represented.

Analyzing the quality of fifteen geolocation models and two baselines, using ten different evaluation measures over four geographic granularities, our study proposes a guide for the evaluation of Twitter user geolocation through the following contributions:

\begin{itemize}
\item We standardize the evaluation process for models to ensure the fairness of comparison. We demonstrate that some older models that were previously thought to be uncompetitive perform comparably to recent approaches.

\item We examine the influence of social media population bias on the quality of geolocation prediction. In particular, we find that a wide range of metrics and a majority class baseline should be used for the evaluation of more complex geolocation models.

\item We assess the effectiveness of current evaluation metrics using rank correlations. We demonstrate that the ranking of user geolocation systems varies based on the evaluation metric and geographic granularity. In some cases, some of the most common evaluation metrics are redundant and should not be used simultaneously.

\item We validate the effectiveness of the proclaimed state-of-the-art geolocation systems using statistical significance testing. We propose a suite of statistical significance tests suitable for the task at hand, based on the employed metric.

\item We study the degree to which metrics can lead to contradictory, yet statistically significant results, concluding that systems should be evaluated using a range of measures.
\end{itemize}

This paper builds upon our own previously-published work~\cite{mourad2018well} with more statistical analysis (effectiveness and significance) through the last three contributions. Our results demonstrate the different properties of measures, which can in turn lead to a better understanding of the differences between models, and to better decision-making based on specific application requirements.

\section{Related Work}

\citet{zheng2018survey} surveyed previous research on the geolocation of Twitter users. They reviewed and summarized all geolocation methods and evaluation metrics employed from an empirical perspective. In this work, we present the metrics from two different perspectives. We briefly introduce the different approaches of inferring a user's location in \S\ref{sec:geomethods}. In \S\ref{sec:evalevol}, we discuss how evaluation metrics for Twitter user geolocation evolved over time. This presentation explains the original intuition behind each metric, reveals the decisions taken by subsequent researchers and the impact of such choices on the evaluation process. We also chart the limitations and commonality of each metric. Examination of biases in social media are detailed in \S\ref{sec:bias}. In \S\ref{sec:evalcomp}, we survey efforts to standardize the evaluation process of Twitter user geolocation and assess the effectiveness of the evaluation metrics employed.

\subsection{Geolocation Methods}
\label{sec:geomethods}

Previous research inferred the location of a Twitter user from different sources of information, namely tweet-text, user's social-network (e.g. followers, following, mentions) and meta-data (e.g. profile location, tweet timezone). Most geolocation methods rely on the first two sources and hence are known in research as text-based and network-based approaches. Text-based methods tend to address geolocation inference as a classification task. They rely on identifying location indicative words (aka local words) over a predefined set of locations (e.g. administrative regions or grid cells). Location sparsity is, therefore, a limitation of text-based approach. The intuition behind network-based methods is that a user is geographically close to their friends. However, if a user is not covered in the training network, a geolocation model will not be able to infer their location. Hence, recent research is focusing on a hybrid approach which combines both approaches.

\citet{jurgens2015geolocation} constructed a benchmark for the network-based approach. They re-implemented the state-of-the-art models, back at the time, and made it publicly available to the research community. In the process to do that, they constructed their own dataset to train the models and ensure fairness of comparison. Recent research, however, still prefer to reconstruct benchmark datasets which were created by the text-based research to evaluate their models, than constructing their own datasets and retraining the available models. We, therefore, choose to focus on text-based approaches to set a reliable benchmark process for the task of Twitter user geolocation regardless of the underlying approach. We highlight the pitfalls of reconstructing Twitter datasets and comparing to results reported in previous research.

\subsection{Geolocation Metric Evolution}
\label{sec:evalevol}

Table \ref{tbl:evaloverview} details a chronological ordering of Geolocation Metrics, which we initially overview and then describe in more detail.

\subsubsection{Overview}
Evaluation of geolocation models was initially measured using \emph{Median} and \emph{Mean} error distances between an estimated and true location \cite{eisenstein2010latent}. The researchers also used accuracy (\emph{Acc}) at the level of states (49) and regions (4). The choice of spatial granularity was influenced by the use of ground truth datasets, which were drawn from the US (the country with the majority of Twitter users in 2010), and for a better interpretability of the results compared to error distance.

Several metrics based on accuracy and/or error distance were introduced. \citet{backstrom2010find} evaluated performance based on the fraction of predictions within $x$ kilometers from the true location using a Cumulative Distribution Function (\emph{\textsc{CDF}}) for all values of $x$ within 10,000 km. Accuracy within $x$ miles from the original city was introduced by \citet{cheng2010you}, as was accuracy within the top $k$ cities (\emph{Acc@k}), and at the level of country by~\cite{hecht2011tweets}. \citet{priedhorsky2014inferring} introduced three new metrics, based on the error distance and the probability of estimation.

\citet{rodrigues2016exploring} reported precision and recall at the level of each city and an overall macro-F1 metric, which was further extended by \citet{mourad2017language} to consider micro, weighted, and macro averaging techniques at the level of the three metrics. Other research employed a combination of these measures, as described in Table \ref{tbl:evaloverview}. Given that \emph{Acc@k}~\cite{cheng2010you} and the metrics introduced by~\citet{priedhorsky2014inferring} were employed only in their respective research, they were not presented in the table.

\begin{sidewaystable}
\caption{An overview of past work. Precision, recall and f1-score are combined in the column PRF. For datasets, names in bold represent the original dataset, empty \#Users and \#Tweets cells means the size of the reconstructed dataset was not reported in the respective work, and Scope refers to the geographical coverage. For testset, percent is the percentage of users in the testset to the whole collection; \#Tpu is the minimum number of tweets per test user.}
\label{tbl:evaloverview}

\begin{center}
\resizebox{\textwidth}{!}{
\begin{tabular}{l | c | c | c | c | c | c | c | c | c | c | c}
	\toprule
	\multirow{2}{*}{} & \multicolumn{5}{c|}{\bf Evaluation Metrics} & \multicolumn{4}{c|}{\bf Datasets} & \multicolumn{2}{c}{\bf Testset} \\
	\cmidrule{2-12}
	& \bf Acc & \bf Acc@161 & \bf Median & \bf Mean & \bf PRF & \bf Name & \bf \#Users & \bf \#Tweets & \bf Scope & \bf \#Users & \bf \#Tpu \\
	\hline
	\citet{eisenstein2010latent} & \checkmark &  & \checkmark & \checkmark &  & \bf GeoText & 9.5k & 380k & US & 1.9k (20\%) & \\
    \hline
	\citet{backstrom2010find} &  & \textsc{CDF} &  &  &  & \bf \textsc{Backstrom} &  &  & US &  & \\
    \hline
	\citet{cheng2010you} & \checkmark & 0--4k &  & \checkmark &  & \bf Cheng & 135k & 4M & US & 5k (3.7\%) & 1000+\\
    \hline
    \citet{wing2011simple} &  &  & \checkmark & \checkmark &  & GeoText &  &  & US &  & \\
    \hline
	\multirow{2}{*}{\citet{roller2012supervised}} &  &  & \checkmark & \checkmark &  & GeoText &  &  & US &  & \\
	&  & \checkmark & \checkmark & \checkmark &  & \bf UTGeo & 449k & 38M & Nth Am & 10k (2.22\%) & \\
    \hline
	\citet{ahmed2013hierarchical} &  &  &  & \checkmark &  & GeoText &  &  & US &  & \\
    \hline
	\multirow{2}{*}{\citet{han2014text}} & \multirow{2}{*}{\checkmark} & \multirow{2}{*}{\checkmark} & \multirow{2}{*}{\checkmark} &  &  & UTGeo &  &  & Nth Am &  & \\
	 & & & &  &  & \bf \textsc{World} & 1.4M & 12M & Global & 10k (0.71\%) & 10+\\
     \hline
	\multirow{2}{*}{\citet{wing2014hierarchical}} &  & \multirow{2}{*}{\checkmark} & \multirow{2}{*}{\checkmark} & \multirow{2}{*}{\checkmark} &  & UTGeo &  &  & Nth Am &  & \\
	 &  &  &  &  &  & \textsc{World} &  &  & Global &  & 10+\\
    \hline
	\citet{priedhorsky2014inferring} &  &  & \checkmark & \checkmark &  & GeoText & 9.5k & 380k & US & 1.9k (20\%) & \\
	\hline
    \citet{jurgens2015geolocation} &  & \textsc{Auc} & \checkmark &  &  & \bf Jurgens &  &  &  &  & \\
    \hline
	\citet{rodrigues2016exploring} & \checkmark & & & & \checkmark & \bf Rodrigues & 11.8k & & Brazil & & \\
	\hline
	\citet{han2016twitter} [W-NUT] & \checkmark & & \checkmark & \checkmark & & \textsc{World} & 1.4M & 12M & Global & 10k (0.71\%) & 10+\\
    \hline
	\multirow{2}{*}{\citet{rahimi2016pigeo,rahimi2017continuous,rahimi2018semi}} & & \multirow{2}{*}{\checkmark} & \multirow{2}{*}{\checkmark} & \multirow{2}{*}{\checkmark} & & UTGeo & & & Nth Am & 10k & \\
	& & & & & & \textsc{World} & 1.4M & 12M & Global & 10k (0.71\%) & 10+ \\
    \hline
	\multirow{2}{*}{\citet{miura2017unifying}} & \multirow{2}{*}{\checkmark} & \multirow{2}{*}{\checkmark} & \multirow{2}{*}{\checkmark} & \multirow{2}{*}{\checkmark} & & UTGeo & 279k & 23.8M & Nth Am & 10k & \\
	& & & & & & \textsc{World} & 782k & 9.03M & Global & 10k & 10+\\
	\hline
	\multirow{2}{*}{\citet{mourad2017language}} & \multirow{2}{*}{\checkmark} & \multirow{2}{*}{\checkmark} & \multirow{2}{*}{\checkmark} & \multirow{2}{*}{\checkmark} & \multirow{2}{*}{\checkmark} & WORLD & 947k & & Global & & \\
	& & & & & & \bf TwArchive & 1.5M & & Global & & \\
	\hline
	\multirow{3}{*}{\citet{do2017multiview}} & & \multirow{3}{*}{\checkmark} & \multirow{3}{*}{\checkmark} & \multirow{3}{*}{\checkmark} & & GeoText & 9.5k & \textgreater370k & US & 1.9k (20\%) & \\
	& & & & & & UTGeo & 450k & 38M & Nth Am & 10k & \\
	& & & & & & WORLD & 1.4M & 12M & Global & 10k & \\
	\hline
	\multirow{3}{*}{\citet{ebrahimi2018twitter}} & & \multirow{3}{*}{\checkmark} & \multirow{3}{*}{\checkmark} & \multirow{3}{*}{\checkmark} & & GeoText & 9.5k & 380k & US & 1.9k (20\%) & \\
	& & & & & & UTGeo & 450k & 38M & Nth Am & 10k & \\
	& & & & & & WORLD & 1.4M & 12M & Global & 10k & \\
	\bottomrule
\end{tabular}
}
\end{center}
\end{sidewaystable}

\subsubsection{Accuracy Error}
\citet{cheng2010you} showed empirically that 30\% of users are placed within 10 miles of their true location, and 51\% within 100 miles after exploring a range from 0 to 4,000 miles.

Subsequent research used the (perhaps) arbitrarily chosen range of 100 miles (161 km) to measure accuracy (\emph{Acc@161})~\cite{roller2012supervised,han2014text,wing2014hierarchical}. Note, the variance in accuracy with respect to the range was tested on a dataset limited to the \textsc{US}. Using a population-based global earth representation,~\footnote{https://github.com/tq010or/acl2013} the average distance between cities and their neighbours was found to be in the range of 32--46 miles~\cite{mourad2017language}, less than half the 100 miles threshold. A system which predicts the location of a user two cities away from his/her home location could be as accurate as a system which predicted the location one city away from the true location. This choice of the tolerance distance questions the appropriateness of \emph{Acc@161} as a measure that suits global geographic models. A somewhere arbitrary threshold is also found in the metric \emph{\textsc{Auc}}, introduced by \citet{jurgens2015geolocation}, which quantifies the graph generated by a \textsc{CDF} into a single number. This number is generated using the range value of 10,000km.

Error distance measures (Mean and Median) can be more accurate than \emph{Acc@161} because they are measured based on the raw estimations of geolocation models without any approximation (discretization, e.g. map to a region such as a city or country). However, they can exhibit a large variability on the measured results and limit evaluation at multiple levels of geographic granularity, which is required by some geolocation applications as mentioned in \S\ref{sec:intro}. On the other hand, metrics based on accuracy and error distance (e.g. \emph{Acc@161}, \emph{\textsc{CDF}}, and \emph{\textsc{AUC}}) strongly depend on the distance thresholds that are selected.

\subsubsection{Dataset Availability}
Table~\ref{tbl:evaloverview} (columns \#Users and \#Tweets) shows a large disparity in the sizes of test datasets. Although Twitter provides access to the public data generated by users, the terms of service limits the sharing of this data to only tweet IDs. Any attempt to reconstruct a dataset used in previous research will be subject to decay, i.e. some tweets will disappear because they have been deleted. In an effort to solve this issue, two approaches were proposed.

First, \citet{jurgens2015everyone} proposed an evaluation framework where the dataset is hosted by a single operator. An experimenter submits a request to the host along with a code. However, the cost to the host of maintaining this service, the difficulty of the development process for the experimenter, and the unprotected intellectual property---the ownership of the code---meant this proposal was not taken up.

Second, \citet{han2016twitter} provided a dataset of tweet IDs for a user geotagging shared task (named \textsc{World}). However, one of the participant teams pointed out that the re-constructed dataset was missing $\sim$25\% of the data~\cite{jayasinghe2016csiro}. Systems are therefore highly likely to be trained on different datasets, based on the time they were re-constructed. Subsequent research~\cite{miura2017unifying} highlighted the same issue using two different datasets (UTGeo and \textsc{World}).

\subsubsection{Earth-representation}
The importance of measures was illustrated when two different models were each found to perform better using different reverse-geocoding technique. \citet{han2014text} demonstrated that a multinomial na\"ive bayes model with feature selection performs better than logistic regression~\cite{roller2012supervised} using city-based representation, \citet{wing2014hierarchical} demonstrated the opposite using uniform grids. 

\subsection{Underlying bias}
\label{sec:bias}

Social media is known to have substantial population biases~\cite{mislove2011understanding}. They relied on the US census data to reveal the sampling bias in Twitter data based on the demographics of Twitter users, namely geographic distribution, gender and race/ethnicity. Not many researchers explored the impact of this bias on either determining the most effective models or evaluation metrics. Focusing on the geographic bias over the urban-rural spectrum, \cite{hecht2014tale} explored three of the most common sources for geotagged information, --- Twitter, Flickr and Foursquare. They showed that there is a population bias towards urban regions.

The first attempt to assess the influence of population bias on the existing models --- Twitter user geolocation --- was done by~\cite{pavalanathan2015confounds}. They explored the influence of Twitter user demographics --- gender and age --- on the tasks of detecting lexical variation over geographic regions and text-based Twitter user geolocation, yet relying on accuracy per category (e.g. male vs female) for evaluation. \citet{johnson2017effect} further explored the impact of geographic bias on the latter task. They differentiated between population bias, and structural bias introduced by algorithmic design. To assess the impact of each of these biases, they explored different sampling techniques on a US rural-urban county based dataset. They demonstrated that existing geolocation approaches perform significantly worse for rural areas than for urban.

Relying on an external gazetteer (US census data, which might not be available for other countries), consolidating geographic regions into two classes only (rural-vs-urban) and finally evaluation of individual categories (e.g. accuracy of male vs female) limits the scalability of the analysis. Most of the recent work relies on datasets with global geographic coverage, with hundreds and thousands of classes, and ignores the existence of biases while designing or evaluating their models. We, therefore, believe that focusing on an enhanced and scalable evaluation metrics (macro averaging in specific) should come first; to reveal such biases and assess their impact on the design of geolocation algorithms.

\subsection{Comparing Geolocation Evaluation Metrics}
\label{sec:evalcomp}

Studies have analyzed the effectiveness of evaluation metrics of Twitter user geolocation.
\citet{jurgens2015geolocation} conducted a comparative analysis of nine geolocation models using a standardized evaluation framework. Their evaluation was limited to a network-based geolocation approach using error distance measures (\textsc{Auc} and Median) and a network specific measure, which does not generalize to other approaches, such as the widely-used text-based ones. More recent work by~\citet{mourad2017language} pointed out that accuracy measures are biased towards locations with a large population. Although they employed a wide range of metrics, their work was limited to a single geolocation model while focusing on the influence of language rather than the effectiveness of the evaluation measures.

In this paper, we focus on the effectiveness of geolocation evaluation regardless of the underlying geolocation approach or the language of text, which entails generalization challenges discussed in the next section. We evaluate the relative performance of fifteen geolocation models and two baselines using all the metrics in Table~\ref{tbl:evaloverview}.

\section{Standardized Evaluation}
\label{sec:stndrdeval}

In considering how to build a standardized evaluation, first, alternate metrics are described that address data imbalance. Second, we examine significance tests to assess the statistical differences between the geolocation models under study. Finally, a unified output format and reverse-geocoding method are employed to assure the fairness of comparisons.

\subsection{Evaluation Metrics}

Much past research treated the problem of geolocating Twitter users as a categorization task. Given the global geographic coverage of such a task (typically thousands of locations), there is an inherent imbalance in the distribution of users over locations. \emph{Acc} and \emph{Acc@161} are biased towards regions with a high population (the majority classes)~\cite{johnson2017effect}. Hence, we investigate conventional measures for multi-class categorization~\cite{sebastiani2002machine,sokolova2009systematic}, which were included partially by~\citet{rodrigues2016exploring} and fully by~\citet{mourad2017language} in the context of Twitter user geolocation. We consider Precision (P), Recall (R) and F1-score (F1) using Micro ($\mu$) and Macro ($M$) averaging. \emph{Precision} is more favored in situations such as when journalists are looking for eyewitnesses within a specific city~\cite{diakopoulos2012finding}. \emph{Recall} is favored in situations such as when these journalists want to increase the search pool~\cite{starbird2012learning}. Both scenarios focus on a single location, where comparison at the micro and macro levels is essential.

Evaluation metrics are categorized into three groups. Continuous evaluation is based on the estimated GPS coordinates ($p$) of a user ($u$), and represented by median and mean error distances from the original gps-point ($\hat{p}$). Discrete evaluation is based on the resolved locations of a user ($l$ and $\hat{l}$ are the predicted and true locations, respectively), and represented by accuracy, precision, recall, and f1-score using micro and macro averaging. Mixed evaluation is based on a combination of continuous and discrete metrics, and represented by accuracy within 100 miles of the true location.

The evaluation metrics considered in this study are defined as:
\subsubsection*{Continuous evaluation}
\begin{align*}
    Error Distance(u) &= great\_circle\{\hat{p}, p\}\\
    Median &= median_{i=0}^{n_\text{users}-1}\{Error Distance(u_i)\}\\
    Mean &= \frac{1}{n_\text{users}} \sum_{i=0}^{n_\text{users}-1}\{Error Distance(u_i)\}
\end{align*}

\subsubsection*{Discrete evaluation}
\begin{equation*}
    Acc = \frac{1}{n_\text{users}} \sum_{i=0}^{n_\text{users}-1} 1(l_i = \hat{l}_i)
\end{equation*}

\begin{align*}
    P_{M} &= \frac{1}{n_\text{locations}} \sum_{i=0}^{n_\text{locations}-1} P(l=l_i, \hat{l}=l_i)\\
    R_{M} &= \frac{1}{n_\text{locations}} \sum_{i=0}^{n_\text{locations}-1} R(l=l_i, \hat{l}=l_i)\\
    F_{M} &= \frac{1}{n_\text{locations}} \sum_{i=0}^{n_\text{locations}-1} F_\beta(l=l_i, \hat{l}=l_i)
\end{align*}

For micro averaging, all precision, recall and f1-score are identical to accuracy~\cite{pedregosa2011scikit}.

\subsubsection*{Mixed evaluation}
\begin{equation*}
    Acc@161(p, \hat{p}) = \frac{1}{n_\text{users}} \sum_{i=0}^{n_\text{users}-1} 1(Error Distance{u_i} \leq 161km)
\end{equation*}

\subsection{Significance Tests}

\citet{dror2018hitchhiker} highlighted the importance of applying statistical significance tests in the field of Natural Language Processing (NLP) to ensure that the experimental results are not coincidental. Given the range of NLP tasks and effectiveness metrics that can be applied, different statistical tests are needed. Based on the decision tree algorithm provided by~\citet{dror2018hitchhiker} for statistical significance test selection, we choose a combination of parametric (t-test) and sampling-free non-parametric tests (sign test, and Wilcoxon). Given the large size of our dataset, parametric tests are applicable because the test statistic follows the normal distribution, and sampling-free tests are computationally less expensive than sampling-based non-parametric tests.

Although~\citet{dror2018hitchhiker} surveyed a large number of NLP papers on different tasks, they did not consider the category frequencies of the datasets. We, therefore, follow the recommendation of~\citet{yang1999re} who considered the appropriate choice of significance tests to measure the statistical differences between categorization models trained on datasets with skewed category distributions. Two types are considered: \textbf{micro} and \textbf{macro tests}.

The \textbf{micro tests} considered in this study are the micro sign test (\emph{s}) and proportions z-test (\emph{p})~\cite{yang1999re}. The former is a binomial test for comparing two systems, A and B, based on binary decisions for all user/location pairs. The latter is used for measures which are proportions: accuracy, precision, and recall. The z-test computation for precision and recall is based on performance scores using micro averaging.

The \textbf{macro tests} include macro sign test (\emph{S}), macro t-test (\emph{T}), and macro t-test after rank transformation (\emph{T'}, a.k.a Wilcoxon)~\cite{yang1999re}. Macro tests were originally based on F1 scores per category as a unit measure, but we employed them for precision and recall as well. The S-test is a binomial sign test used to compare two systems, A and B, based on the paired F1 values for individual locations. While the S-test reduces the influence of outliers, it may be insensitive in performance comparison because it ignores the magnitude of differences between F1 values. Insensitivity issues are resolved in T by considering the absolute differences between paired F1 values in a relevance t-test. However, T becomes sensitive when F1 values are unstable, specifically for low frequency locations. Finally, the Wilcoxon T' provides a compromise between S-test and T by considering the rank differences between paired F1 values for individual locations.

We use two-sided versions of the tests, as they avoid prior expectation about the direction of the effect and are more conservative.

\subsection{Unified Output and Reverse-Geocoding}

\begin{figure}
\centering
\includegraphics[width=0.5\textwidth]{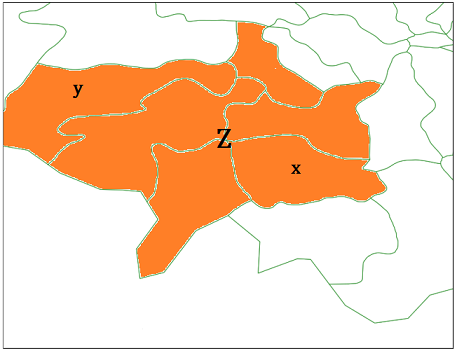}
\caption{Example of unfair comparison between systems with different underlying earth representations. Cell $x$ is the home location of a user and cell $y$ is the predicted location by system $A$. The orange cells represent the home and predicted city ($Z$) of a user by system $B$.}
\label{fig:unifiedoutput}
\end{figure}

When comparing models, it is necessary to train and test on the same dataset and to use models that output the same earth representation. Figure~\ref{fig:unifiedoutput} shows an example of an unfair comparison between models producing different outputs on the same dataset. Assume we have two models: $A$ and $B$. $A$ represents the earth as polygons (green outlined cells) and $B$ represents the earth as cities. A user's home location is identified as polygon $x$ inside city $Z$ (the orange area). Now assume $A$ predicted the location of this user as $y$, and $B$ predicted it as city $Z$. Based on the underlying representation of each model, the prediction of model $A$ will be considered incorrect while the prediction of model $B$ is correct.

In order to avoid such inconsistency, we unified the output of all the models to be GPS coordinates as suggested by~\cite{jurgens2015everyone}. We additionally resolved the coordinates to a location using a single online reverse-geocoding API\footnote{There is a trade-off between replicability, efficiency and cost, when choosing a reverse-geocoding API. An offline reverse-geocoding would be fast, but requires implementation and sharing the code base. On the other hand, online reverse-geocoding is easy to consume, but limited by a specific number of requests per day. Free APIs have a small limit, e.g. 2.5k requests per day for Nominatim, While commercial APIs have a larger limit, e.g. 100k requests per day for Google Reverse-Geocoding API V3. It took two weeks to reverse-geocode our local dataset of the size 1.5M users, using Google API.} before evaluation. Using a single reverse-geocoding API not only guarantees a fair comparison over the same set of locations (classes), it also allows evaluation over different granularities. In this work, we report the model performance at city and country level. We calculated county and state level as well, but trends are consistent.

\section{Experimental Setup}

We examine two sets of systems. The first set (\textsc{Local}) includes four geolocation models and two baselines, trained and tested (over 30k users) locally over the same data collection with free earth representation to evaluate the considered process. The second set (W-NUT) includes eleven submissions from a geolocation shared task, to assess the robustness of our proposed metrics~\cite{han2016twitter}. Although the published results for participating models were evaluated at city level only, we were able to infer output at country level based on information released by W-NUT organizers.

\subsection{\textsc{Local} Models}

\subsubsection{Data Collection Method}
We employed a geographically global geotagged tweet collection, \textbf{TwArchive}, holding content since 2013\footnote{https://archive.org/details/twitterstream\&tab=collection} drawn from the 1\% sample Twitter public API stream. We used a 2014 subset spanning nine months. We focus on English tweets only as identified by langid.py~\cite{lui2012langid}. Non-geotagged and duplicate tweets were removed using user id and tweet text. For the sake of a standard evaluation, users with unresolved home location---based on the model that accepts home locations in the form of cities instead of GPS coordinates~\cite{han2014text}---were removed from the dataset. The total number of users and tweets after pre-processing is $\sim$1.5 million and $\sim$3.1 million respectively.

\subsubsection{Ground Truth}
The home location of a user was identified as the geometric median of their geotagged tweets~\cite{jurgens2013s}. Such a point is the minimum error distance to all locations of a user. The median has been shown to be more accurate in identifying the home location of a user at a finer granularity than other approaches~\cite{poulston2017hyperlocal}. The distance between any two GPS points is measured using the great circle distance method.

\subsubsection{Geolocation Inference Models}
Four models and two baselines were compared using four classification methods and two statistical methods. The models were chosen based on their availability, reproducibility, and recency.

\emph{\citet{roller2012supervised}} \textsc{(Rl12)} proposed an adaptive grid-based representation with a trained probabilistic language model per cell. Each cell has the same number of users, but a different geographical area. We employ their best reported parameter values for constructing the grid to retrain their model~\footnote{https://github.com/utcompling/textgrounder/wiki/RollerEtAl\_EMNLP2012} on our local dataset. The output represents the centroid of the predicted cell.

\emph{\citet{han2014text}} \textsc{(Hn14)} locates users to one of 3,709 cities. We re-implemented their system, focusing on the part that uses Location Indicative Words (LIW) drawn from tweets, where mainstream noisy words were filtered out using their best reported feature selection method, Information Gain Ratio. The output represents the centre of the predicted city.

\emph{\citet{rahimi2016pigeo}} \textsc{(Rm16)} assigns a user to one of 930 non-overlapping geographic clusters based on the similarity of content. Their geotagging tool, Pigeo~\footnote{https://github.com/afshinrahimi/pigeo}, allowed retraining their text-based model on our local dataset. The output represents the median of the predicted cluster.

\emph{Linear SVM} \textsc{(Lsvm)} is a classic approach for imbalanced learning unlike Na\"ive Bayes. It is a variation of \textsc{Hn14} by replacing the classifier. The linear kernel is known to perform well over large datasets within a reasonable time.

\emph{Majority Class} \textsc{(Mc)} is a baseline that always predicts the most frequent class in the training set. \citet{yang1999evaluation} pointed out that in the case of a low average training instances per category (which applies here) the \emph{majority class trivial classifier} tends to outperform all non-trivial classifiers. It was used as a baseline in previous work~\cite{han2014text,mourad2017language}.

\emph{Stratified Sampling} \textsc{(Ss)} is a baseline which picks a single class randomly biased by the proportion of each class in the training set. \textsc{Ss} is expected to be a strong baseline for a classification task with multiple majority (or close to majority) classes, unlike \textsc{Mc} which originated in binary classification.

Both baselines were implemented using scikit dummy classifier~\cite{pedregosa2011scikit} and output a class, not a GPS coordinate. Measures that require a GPS coordinate to measure distance, Acc@161 and mean/median error, were consequently not used to evaluate the baselines.

\subsection{W-NUT Models}

W-NUT\footnote{https://noisy-text.github.io/2016/geo-shared-task.html} is a shared task for predicting the location of posts and users from a pre-defined set of cities~\cite{han2016twitter}. We analyze the results of eleven systems in the user geolocation prediction task (submitted by five teams). The top two submissions were based on ensemble learning (\textsc{Csiro.1}) and neural networks (\textsc{FujiXerox.2}), making use of multiple sources of information, including tweets, user self-declared location, timezone values, and other features. One submission used tweet text only (\textsc{Ibm}). Two teams (\textsc{Aist} and \textsc{Drexel}) did not submit a description of their submissions.

\section{Results}

Table~\ref{tbl:results1} details the results of our experiments on two sets of systems (\textsc{Local} and W-NUT) across all metrics mentioned in Table~\ref{tbl:evaloverview}; PRF (precision, recall, f1-score) are calculated using $\mu$ and $M$ averaging; using the output levels city and country. Error distance metrics (Median and Mean) are measured between the home and estimated GPS coordinates of a user. The best scoring systems for each metric are highlighted in bold.

We first compare which systems are judged best under different evaluations, next we examine rank correlations of systems, and finally study significant differences. For each experiment, we compare across output levels (i.e. city vs country) and at the same output level (i.e. city or country).

\begin{sidewaystable}
\caption{Evaluation based on all metrics at the level of city and country and sorted in a descending order of Acc.}
\label{tbl:results1}

\centering
\resizebox{\textwidth}{!}{%
\begin{tabular}{cl cc|lll|lll | cc|lll|lll | rr}
\toprule
&                & \multicolumn{8}{c|}{City} & \multicolumn{8}{c|}{Country} & \multirow{2}{*}{Median} & \multirow{2}{*}{Mean} \\
&                        & Acc   & Acc@161 & $P_{\mu}$ & $R_{\mu}$ & $F1_{\mu}$ & $P_{M}$ & $R_{M}$ & $F1_{M}$ & Acc   & Acc@161 & $P_{\mu}$ & $R_{\mu}$ & $F1_{\mu}$ & $P_{M}$ & $R_{M}$ & $F1_{M}$ &      & \\
\midrule
\multirow{5}{*}{\rotatebox[origin=c]{90}{\textsc{Local}}} & \textsc{Lsvm}          & \bf 0.145 & 0.193     & 0.085     & 0.068     & \bf 0.075 & 0.045     & \bf 0.040 & \bf 0.039 & 0.446     & 0.448     & 0.447     & 0.446     & 0.447     & 0.098     & 0.113     & 0.099     & 3656     & 5936     \\
& \textsc{Rl12}          & 0.128     & \bf 0.228 & \bf 0.114 & 0.050     & 0.070     & 0.036     & 0.020     & 0.023     & \bf 0.615 & \bf 0.619 & \bf 0.621 & \bf 0.615 & \bf 0.618 & 0.144     & \bf 0.138 & \bf 0.133 & \bf 1740 & \bf 3785 \\
& \textsc{Hn14}          & 0.127     & 0.182     & 0.068     & \bf 0.070 & 0.069     & \bf 0.091 & 0.014     & 0.020     & 0.599     & 0.600     & 0.600     & 0.600     & 0.600     & \bf 0.241 & 0.050     & 0.068     & 3128     & 4489     \\
& \textsc{Rm16}          & 0.074     & 0.132     & 0.030     & 0.021     & 0.025     & 0.007     & 0.001     & 0.001     & 0.315     & 0.316     & 0.315     & 0.315     & 0.315     & 0.062     & 0.015     & 0.015     & 5909     & 5653     \\
& \textsc{Mc}            & 0.018     & 0.000     & 0.018     & 0.020     & 0.019     & 0.000     & 0.000     & 0.000     & 0.523     & 0.000     & 0.523     & 0.524     & 0.523     & 0.004     & 0.007     & 0.005     & ---      & ---      \\
& \textsc{Ss}            & 0.002     & 0.000     & 0.003     & 0.002     & 0.002     & 0.001     & 0.000     & 0.000     & 0.301     & 0.000     & 0.302     & 0.302     & 0.302     & 0.007     & 0.007     & 0.007     & ---      & ---      \\
\midrule
\multirow{10}{*}{\rotatebox[origin=c]{90}{W-NUT}} & \textsc{Csiro.1}      & \bf 0.529 & 0.636  & \bf 0.544 & \bf 0.529 & \bf 0.537 & 0.545     & 0.432     & 0.454     & 0.798     & 0.799     & 0.798     & 0.798     & 0.798     & 0.661     & \bf 0.538 & \bf 0.568 & 21       & 1928     \\
& \textsc{Csiro.2}      & 0.523     & 0.619      & \bf 0.544     & 0.523     & 0.533     & 0.555     & \bf 0.434 & \bf 0.458 & 0.787     & 0.789     & 0.788     & 0.787     & 0.787     & 0.653     & 0.535     & 0.561     & 23       & 2071     \\
& \textsc{Csiro.3}      & 0.503     & 0.585      & 0.529     & 0.503     & 0.516     & \bf 0.576 & 0.422     & 0.455     & 0.771     & 0.773     & 0.772     & 0.771     & 0.771     & 0.662     & 0.530     & 0.560     & 30       & 2242     \\
& \textsc{FujiXerox.2}  & 0.476     & 0.635      & 0.481     & 0.476     & 0.478     & 0.358     & 0.279     & 0.289     & 0.866     & 0.868     & 0.866     & 0.866     & 0.866     & \bf 0.692 & 0.519     & 0.562     & \bf 16   & 1122     \\
& \textsc{FujiXerox.1}  & 0.464     & \bf 0.645  & 0.468     & 0.464     & 0.466     & 0.313     & 0.253     & 0.253     & \bf 0.883 & \bf 0.886 & \bf 0.884 & \bf 0.883 & \bf 0.884 & 0.634     & 0.514     & 0.542     & 20       & \bf 963  \\
& \textsc{FujiXerox.3}  & 0.452     & 0.629      & 0.455     & 0.452     & 0.453     & 0.283     & 0.243     & 0.237     & 0.869     & 0.872     & 0.869     & 0.869     & 0.869     & 0.621     & 0.502     & 0.527     & 28       & 1084     \\
& \textsc{Drexel.3}     & 0.352     & 0.474      & 0.367     & 0.352     & 0.359     & 0.348     & 0.230     & 0.253     & 0.686     & 0.689     & 0.701     & 0.686     & 0.693     & 0.631     & 0.494     & 0.530     & 262      & 3124     \\
& \textsc{Ibm.1}        & 0.225     & 0.349      & 0.225     & 0.225     & 0.225     & 0.099     & 0.049     & 0.053     & 0.706     & 0.707     & 0.706     & 0.706     & 0.706     & 0.306     & 0.148     & 0.169     & 630      & 2860     \\
& \textsc{Aist.1}       & 0.098     & 0.199      & 0.103     & 0.098     & 0.100     & 0.123     & 0.052     & 0.063     & 0.562     & 0.564     & 0.565     & 0.562     & 0.564     & 0.297     & 0.107     & 0.137     & 1711     & 4002     \\
& \textsc{Drexel.1}     & 0.080     & 0.140      & 0.082     & 0.080     & 0.081     & 0.062     & 0.025     & 0.031     & 0.354     & 0.355     & 0.355     & 0.354     & 0.355     & 0.157     & 0.072     & 0.086     & 5714     & 6053     \\
& \textsc{Drexel.2}     & 0.079     & 0.135      & 0.082     & 0.079     & 0.080     & 0.056     & 0.024     & 0.029     & 0.435     & 0.435     & 0.443     & 0.435     & 0.439     & 0.168     & 0.072     & 0.090     & 4000     & 6161     \\
\bottomrule
\end{tabular}
}
\end{sidewaystable}

\subsection{Best system}
We compare two forms of evaluation based on metric commonality as shown in Table~\ref{tbl:evaloverview}: most common metrics (Acc, Acc@161, Median and Mean error distances) and alternate metrics (PRF using $\mu$ vs $M$ averaging).

\subsubsection{Unified output influence using most-common metrics}
\label{sec:unified}

\begin{figure}
\centering
\includegraphics[width=\textwidth]{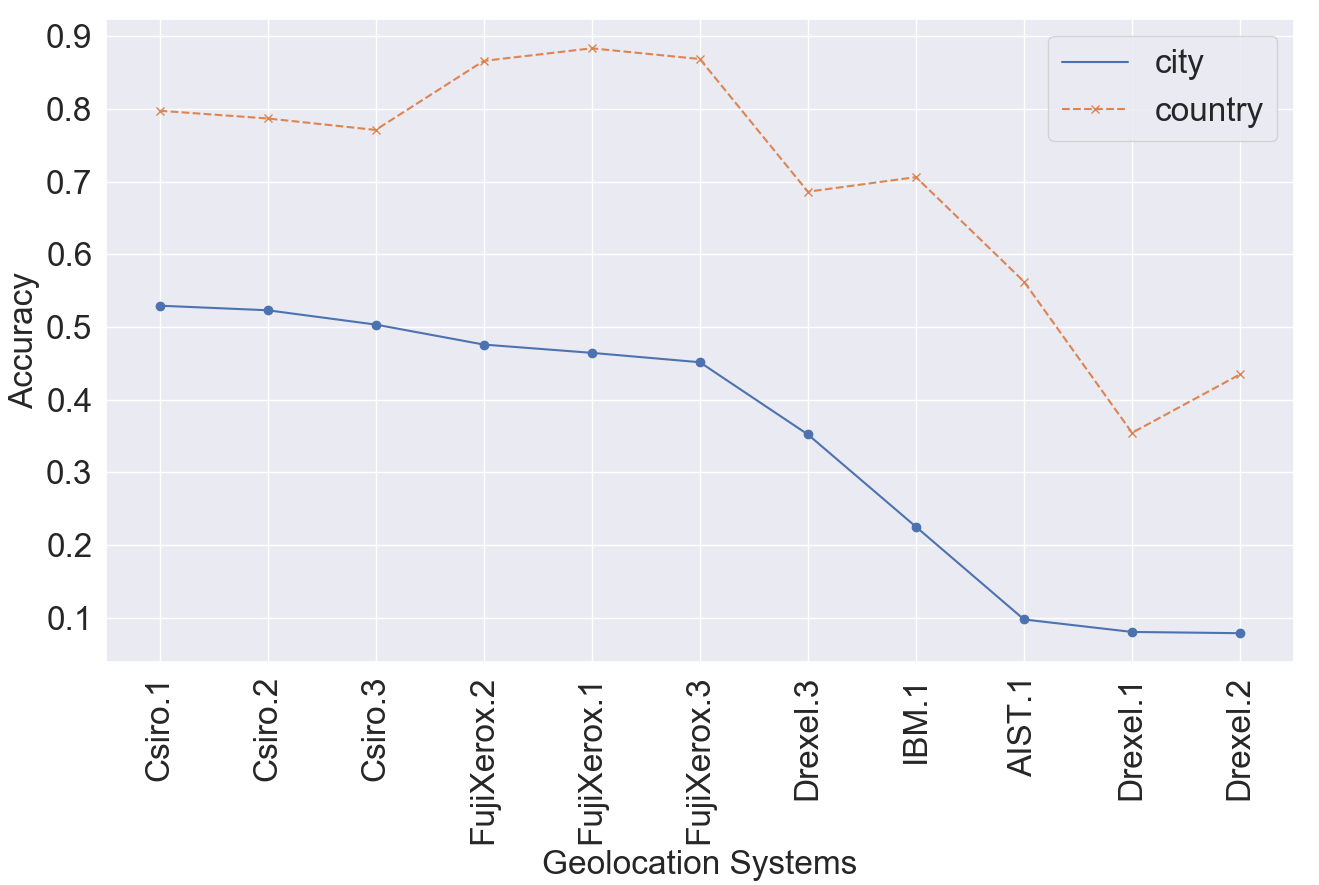}
\caption{Evaluation of W-NUT based-on accuracy at the levels of city and country, ordered by city in a descending order.}
\label{fig:acc-wnut}
\end{figure}

The country and city representations are evaluated using two measures: Acc and Acc@161, which report different best performing geolocation models in the \textsc{Local} and W-NUT sets at the city level, respectively. In terms of accuracy measures, results in the \textsc{Local} section of Table~\ref{tbl:results1} show that \textsc{Rl12} and \textsc{Hn14} are competitive in terms of Acc at the level of city, while \textsc{Rl12} achieves better results in terms of Acc at the level of country and Acc@161 at both levels. On the other hand, the \textsc{Lsvm} model achieves the best Acc at the level of city only \ms{What did past research say about Lsvm compared to the Rl12 and Hn14 systems?}\am{Nothing. After going through literature again, I found that it wasn't considered in the first place. So, what's now? :-) I don't think we should emphasize on this, not a big contribution.}.

To further illustrate the differences found when using city and country representations, the W-NUT systems, measured using Acc, are shown in Figure~\ref{fig:acc-wnut}. Standardization enables the comparison of the best performance of each geolocation model.

\begin{figure}
\centering
\includegraphics[width=\textwidth]{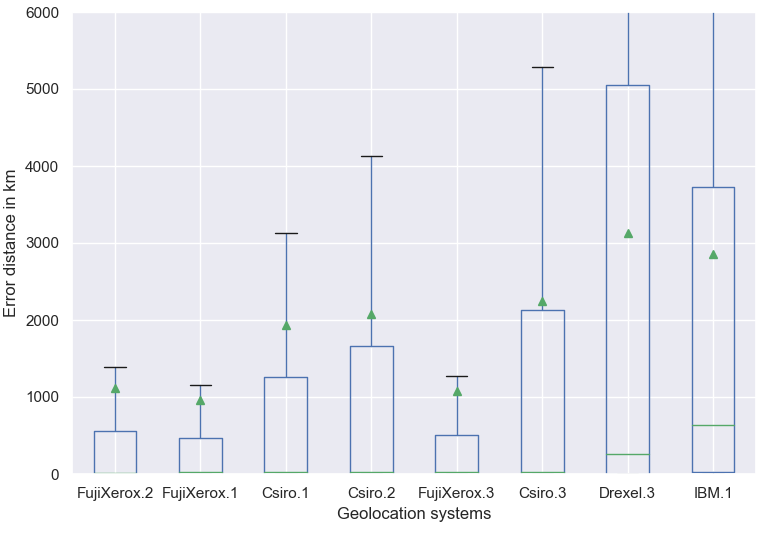}
\caption{Evaluation of W-NUT based-on error distance metrics (Median and Mean) in km.}
\label{fig:errdist-wnut}
\end{figure}

We examine the error distance measures to try to understand the observed differences in best \textsc{Local} system. There is a gap in performance between the grid based model (\textsc{Rl12}) and the city (\textsc{Hn14} and \textsc{Lsvm}) or region/cluster (\textsc{Rm16}) based models, see Table~\ref{tbl:results1}. This gap is related to the geographic footprint per unit of the underlying earth representation. Grid-based approaches tend to have the lowest error distances (because they are calculated from the center of the predicted cell), followed by city-based, and finally region-based approaches.

For the W-NUT shared task, we observe that the \textsc{FujiXerox} submissions tend to have slightly better Acc@161 at the level of city than the \textsc{Csiro} submissions (see Table~\ref{tbl:results1}). At the level of country, however, the \textsc{FujiXerox} submissions achieve much better results than \textsc{Csiro}, which is correlated to the gap in the mean error distance (in favor of \textsc{FujiXerox} models) despite having competitive median error distance, as we will show later. Note that the original WNUT shared task did not evaluate the participating systems at the level of country.

The distribution and results for the error distance measures are represented in more detail using a box plot in Figure~\ref{fig:errdist-wnut}. The green triangles represent the mean error distance for each system. An upper threshold distance of 6,000 km was applied and the worst three systems (\textsc{Aist.1}, \textsc{Drexel.1}, and \textsc{Drexel.2}) were excluded so that details can be seen. We can observe the large variance in 50-75\% percentile between \textsc{FujiXerox} submissions and \textsc{Csiro}. Previous research~\cite{han2014text,melo2017automated} promoted the usage of median error distance to evaluate user geolocation because it is more robust to outliers than the mean, and easy to interpret the results in comparison to accuracy. However, the boxplot quantifies the variance in error distance, and 25\% can not be considered as outliers in this case. The mean error distance therefore is a more effective measure than the median in this context. \textsc{FujiXerox} and \textsc{Csiro} submissions have competitive results in terms of the median error distance, while \textsc{FujiXerox} submissions are much better in terms of the mean error distance and have less variance in their estimations.

Results in the \textsc{Local} section of Table~\ref{tbl:results1} show that the two baselines for the locally deployed geolocation models (\textsc{Mc} and \textsc{Ss}) perform poorly at the level of city. In contrast, \textsc{Mc} establishes a strong baseline at the level of country, where it performs much better than \textsc{Rm16}, \textsc{Lsvm} and \textsc{Ss}. \textsc{Mc} is effective at the level of country because of the lower number of countries (few hundreds) compared to cities (few thousands). Given the large size of the training set (1.5 million), the sparsity at the country level will be less, still with bias in the distribution, which also explains why the Na\"ive Bayes based model (\textsc{Hn14}) performs better than \textsc{Lsvm} in this case. The \textsc{Ss} baseline performs poorly, which suggests it should not be considered as a baseline. At this stage, the use of a simple \textsc{Mc} baseline and Acc did not reveal the influence of imbalance as~\cite{yang1999evaluation} suggested. Therefore, we consider evaluation using different averaging techniques and alternative measures to provide a better insight into the influence of imbalance.

\subsubsection{Imbalance Influence using alternate metrics}

The three evaluation measures (PRF) that use the two averaging methods can be compared across city and country giving six $\mu$ vs $M$ comparisons. Across those six, the best system is different in 67\% and 100\% of the comparisons in the \textsc{Local} and W-NUT sets, respectively. \ms{Is this different to what's been shown in the past?}\am{Micro-macro comparison and the city-country contrast were not considered in previous research (except my ecir paper for the former). The main focus was always different, whether it's the lexical variation over geographic granularities or the comparisons of research models at the same granularity.}

A consistent drop in performance can be seen from $\mu$ to $M$, see columns $P_{\mu}$ to $F1_{M}$ of Table~\ref{tbl:results1}. While \textsc{Rl12} and \textsc{Hn14} are competitive at the level of Acc, \textsc{Rl12} tend to have higher precision than \textsc{Hn14} using micro averaging, and vice versa using macro averaging. \textsc{Lsvm} is another example where Acc is a limited measure when comparing to other systems. While \textsc{Lsvm} achieves the best Acc at the level of city, it tends to have less precision than \textsc{Rl12} using micro averaging and \textsc{Hn14} using macro averaging, yet has higher recall achieving the best F1-score among all systems in \textsc{Local}. \textsc{Mc} is still competitive at the country level using micro averaging, achieving higher PRF than \textsc{Rm16} and \textsc{Lsvm}.

If we consider both unified output and imbalance influences, in W-NUT, collectively the \textsc{Csiro} submissions outperform \textsc{FujiXerox} at the level of city across all the evaluation metrics, except for Acc@161 and error distance measures. On the other hand, \textsc{FujiXerox} submissions outperform \textsc{Csiro} at the level of country in terms of accuracy, micro averaging and error distance measures, and vice versa using macro averaging, except for macro precision ($P_{M}$).

To summarize the best system analysis, we demonstrated (in \S\ref{sec:unified}) that unifying the output format and reverse-geocoding locations before evaluation are essential to ensure the fairness of comparison. A majority class baseline is recommended at the country level in the case of using Acc or micro averaging method. The alternate metrics (macro averaging in specific) should be used to evaluate the influence of data imbalance on the quality of geolocation prediction. The question now is: how to quantify the effectiveness of using different evaluation measures?

\subsection{Rank correlations}

\begin{figure}
\begin{subfigure}{\textwidth}
\centering
\includegraphics[scale=0.4]{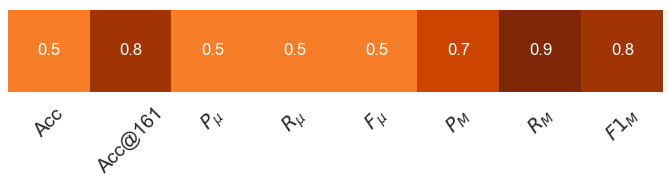}
\caption{Rank correlations across City and Country. Median and Mean error distances are excluded because geographic granularity is not applicable.}
\label{fig:rc-city-country}
\end{subfigure}

\begin{subfigure}{\textwidth}
\centering
\includegraphics[width=0.5\textwidth]{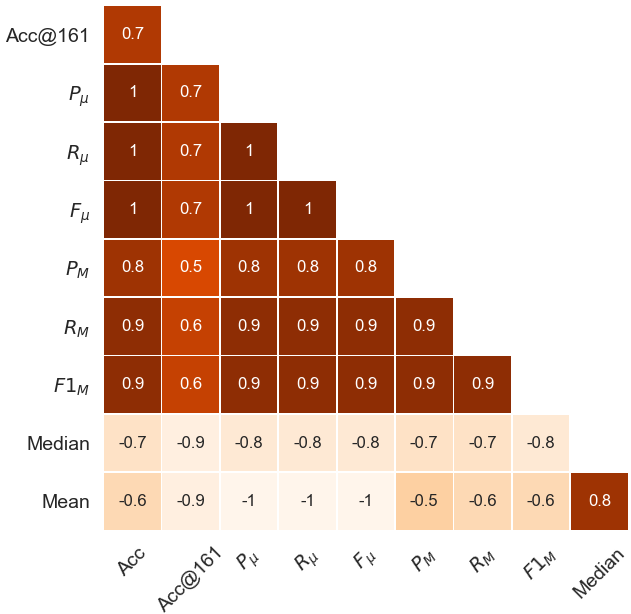}
\caption{Rank correlations at the level of City.}
\label{fig:rc-city}
\end{subfigure}

\begin{subfigure}{\textwidth}
\centering
\includegraphics[width=0.5\textwidth]{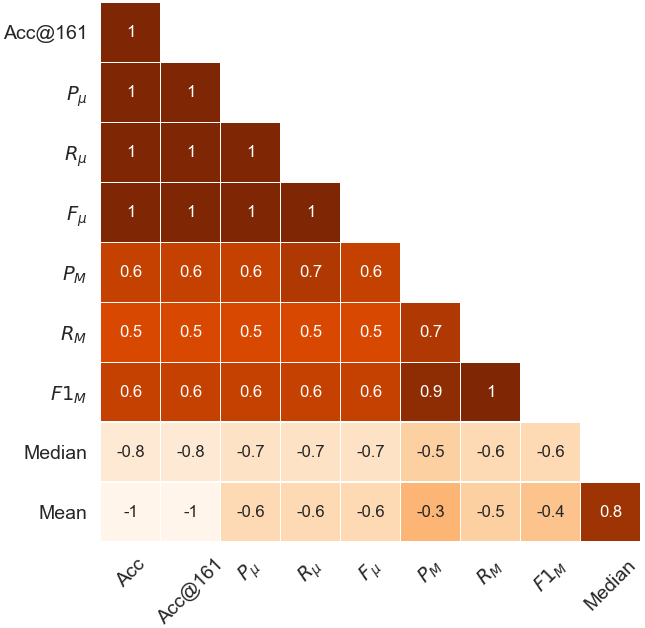}
\caption{Rank correlations at the level of Country.}
\label{fig:rc-country}
\end{subfigure}

\caption{Kendall's $\tau_{B}$ rank correlations between pairs of effectiveness metrics for the W-NUT collection, $p<=0.05$.}
\label{fig:rankcorr}
\end{figure}

Kendall's $\tau$ is a correlation measure that quantifies the agreement between two ranked lists. We calculated $\tau_{B}$ for all combinations of the employed metrics, see Figure~\ref{fig:rankcorr}. Note, because the optimal value for distance metrics is 0 and the optimal value for the other metrics is 1, the optimal correlation between those two is -1; the optimal correlation between the non-distance metrics is 1. As the \textsc{Local} collection only includes four non-baseline systems, the range of $\tau_{B}$ values is limited, we therefore focus our analysis on the W-NUT data.

A strong correlation of any metric across different geographic granularities indicates the consistency of such a measure in ranking geolocation models. On the contrary, a strong correlation between any two metrics at the same level of geographic granularity (e.g. city) indicates less benefit from using both metrics at the same time. Hence, a moderate or weak correlation suggests using both measures is important so a more complete picture of system effectiveness is conveyed.

Considering city vs country (Figure~\ref{fig:rc-city-country}), we observe a weak correlation between ranking models across city and country using the commonly used Acc and micro averaging measures. Using macro averaging measures, a strong correlation exists, similarly for Acc@161. This finding suggests that macro measures and Acc@161 are more robust for comparison across geographic granularities.

Considering micro vs macro at the city level (Figure~\ref{fig:rc-city}), we observe strong correlations across the three micro and macro measures (0.8, 0.9, 0.9). Acc@161, median and mean error distances also have mutual strong correlations. On the other hand, Acc, median and mean error distances have weak correlations. This contrast in correlations, therefore, suggests not relying solely on measures driven from the error distance (Median, Mean, Acc@161) because they depend on the underlying earth representation, i.e. grid-based representation will always achieve better results than city and cluster based representations in terms of these metrics, even if the accuracy of city and cluster based models are better.

Considering micro vs macro at the level of country (Figure~\ref{fig:rc-country}), we observe moderate correlations across the three micro and macro metrics (0.6, 0.5, 0.6). The most common metrics (Acc, Acc@161, Median and Mean) and micro averaging metrics tend to have strong correlations. On the other hand, they tend to have moderate correlations with macro averaging metrics, except for the Median error distance. Therefore, a combination of micro and macro metrics or most common metrics and macro metrics is recommended.

\subsection{Statistical Significance}
As was apparent from the system effectiveness scores in Table~\ref{tbl:results1}, some of the results occurred within a close range. Statistical significance tests are therefore important to establish confidence that differences are not just due to chance.

\begin{figure}
\begin{subfigure}{\textwidth}
\centering
\includegraphics[width=0.65\textwidth]{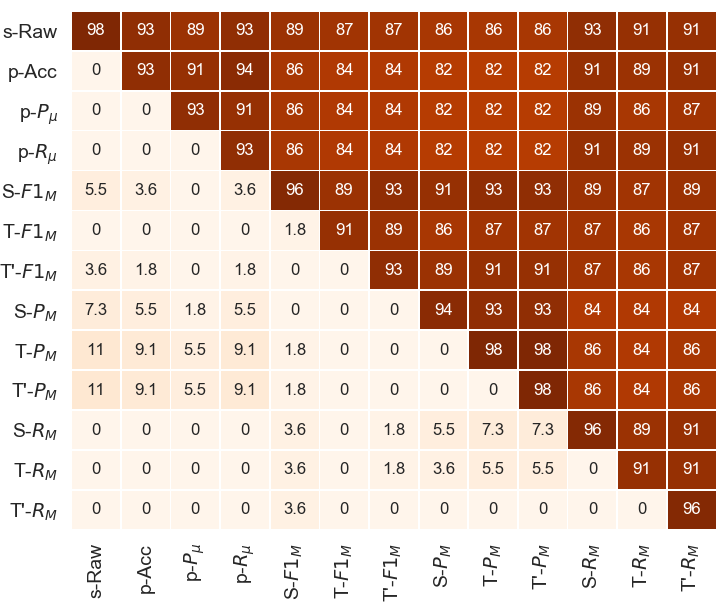}
\caption{City-level}
\label{fig:sad-city}
\end{subfigure}

\begin{subfigure}{\textwidth}
\centering
\includegraphics[width=0.65\textwidth]{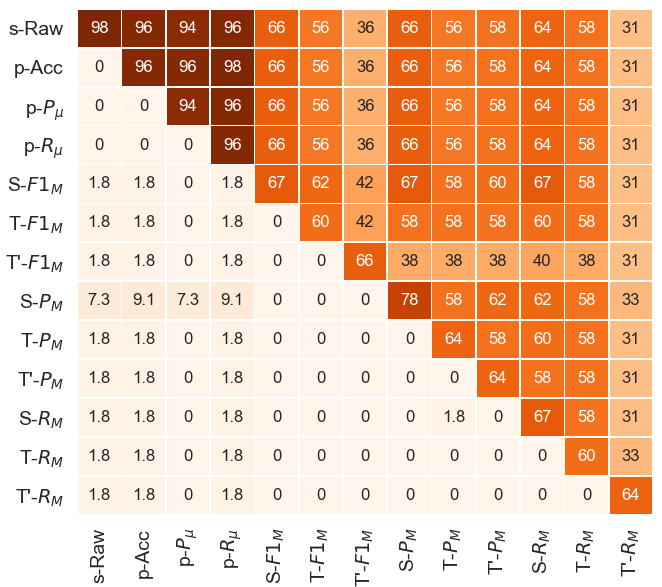}
\caption{Country-level}
\label{fig:sad-country}
\end{subfigure}

\caption{Significant agreements and disagreements, $p=0.05$. W-NUT: 11 systems, 55 system pairs. Micro tests are s-Raw, p-Acc, p-P$_\mu$, and p-R$_\mu$, while the rest represent Macro tests. Significance tests abbreviations stand for: s$\rightarrow$sign-test, p$\rightarrow$proportions z-test, S$\rightarrow$macro sign-test, T$\rightarrow$macro t-test, and T'$\rightarrow$Wilcoxon test.}
\end{figure}

Following~\cite{moffat2012models}, the outcome of a significance test will be categorized into one of two classes. Given two models A and B, calculated for two metrics X and Y, suppose that significance tests are run on the models' outputs using both metrics. If they show statistically significant results for both metrics that system A is better than system B (or vice versa), that would be considered a statistically significant active agreement (SSA). Statistical significant differences, but with contradicting superiority on systems, would be considered a statistically significant active disagreement (SSD).

Figures~\ref{fig:sad-city} and~\ref{fig:sad-country} summarize the results of the statistical significance tests for the W-NUT collection at the city and country levels, respectively. Each figure summarizes the significant agreements and disagreements. The diagonal values represent the percentage of systems pairs that are significantly different based on a single metric (discriminative power), the values above the diagonal show the percentage of SSA, the values below the diagonal show the percentage of SSD. As can be seen, there are many more agreements than disagreements.

Considering city vs country, we observe that the discriminative power of the evaluation metrics (on the diagonal) and the percentage of SSA at the city level (above the diagonal) are always over 80\% (see Figure~\ref{fig:sad-city}). They are much lower at the level of country for all comparisons involving macro tests (31--78\%, see Figure~\ref{fig:sad-country}), which suggests that there is no huge difference in performance between geolocation models. On the contrary, the percentage of SSD (below the diagonal) at the level of country is much lower than at the city level. These results support the importance of using macro metrics for cross granularity evaluation suggested in the previous section.

Considering precision and recall, at the city level, the discriminative power (on the diagonal) using macro averaging is better than micro; macro averaging is able to capture more statistically significant differences for both precision and recall. At the country level, the opposite is true: micro measures are more discriminative than macro. The percentage of SSA involving micro metrics (first fours columns), above the diagonal, are observed to be higher than macro metrics. The percentage of SSA involving macro metrics can drop down to 30.9\%. The level of disagreements (SSD) are generally low or zero. However, the occurrence rate is sometimes as high as 10.9\% (for example, for T-$P_{M}$ and s-Raw at the city level), similarly for tests involving macro metrics. These are cases where experiments would have led to contradictory conclusions about statistically significant differences in system effectiveness, simply based on the metric that was chosen for evaluation. For general evaluation, a macro-micro statistical significance comparison is recommended.

\section{Discussion and Limitations}

Datasets built using Twitter cannot be fully shared and are practically irreproducible because they are subject to decay over time. This challenge will persist, unless Twitter changes their policy and the end-users give their consent to make use of their data. Sharing datasets~\cite{han2016twitter}, therefore, is not feasible. Building centralized frameworks where all researchers submit their systems~\cite{jurgens2015everyone}, isn't practical as well. Hence, every researcher will likely need to create their own datasets, which will normalize the impact of confounding factors, such as data decay, pre-processing, and ground-truth construction. However, this step requires other researchers to share their systems with the ability to retrain their models.

The global geographic coverage of social media means that datasets are naturally imbalanced in terms of locations, with bias towards big cities. Given that classification is a common approach to predict the location of a Twitter user, it is important to highlight the large number of classes (thousands) involved in the learning process. For general evaluation such as in \textsc{Wnut} shared-task~\citet{han2016twitter} or applications treating urban and rural locations with the same degree of importance, a macro versus micro evaluation should be employed to address the limits of the most common metrics (accuracy and error distance). A majority class baseline is also recommended at the level of coarse geographic granularities, state and country in particular, as it achieved competitive results. Finally, we encourage researchers to report the probability of their predictions/estimations, as opposed to binary classification outputs, to allow for assessing the effectiveness of more evaluation metrics, such as \textsc{CDF} (\S~\ref{sec:evalevol}) and \textsc{Auc} (\S~\ref{sec:evalcomp}).

With the large number of explored metrics, Kendall's $\tau$ rank correlation test is recommended to quantify the agreement between pairs of metrics. Our results showed that Acc@161, and macro metrics are more consistent and highly correlated across different granularities in comparison to Acc and micro metrics. We demonstrated that error distance metrics (Median, and Mean) and Acc@161 are dependent on the underlying earth representation. While they are highly correlated at the same geographic granularity, they do not convey different information (redundant) and error distance measures are insensitive to evaluation at several geographic granularities. Hence, they should not be used as sole measures for evaluation, which is still the common practice~\cite{rahimi2018semi,ebrahimi2018twitter,bakerman2018twitter,miura2017unifying}, specially using Acc@161 at fine granularities (city and county). A combination of macro metrics (precision, recall and f1-score) and either micro metrics or accuracy and error metrics are recommended for evaluation.

Statistical significance tests at micro and macro levels were employed to assess the effectiveness of the evaluation metrics at both levels. Using SSA and SSD to summarize the outcome, our results revealed the disparity in agreements and disagreements between tests based on the chosen evaluation metric and geographic granularity. The SSAs between micro and macro tests are higher (better) at the level of city than country. The SSDs are higher (worse) at the level of city than country. To the best of our knowledge, only few recent works applied statistical analysis~\cite{miura2017unifying}, and choosing the right tests can be challenging. Statistical significance testing is essential to draw robust conclusions about the state-of-the-art. In the context of multi-classification and data imbalance, we recommend this list of two-sided tests: 
\begin{enumerate*}[label=\roman*.]
    \item Micro sign test (s) and proportions z-test (p) for micro evaluation using raw predictions, accuracy, precision and recall.
    \item Macro sign test (S), macro t-test (T), and Wilcoxon test for macro evaluation using precision, recall and f1-score.
\end{enumerate*}

The choice of evaluation metrics should be justified by the needs of the applications and the underlying earth representation. A standardized evaluation process, which unified the output format, allowed the comparison of systems with different earth representations. We demonstrated that different systems were found to be best for different underlying representations using an evaluation process including eight measures. Unlike previous research~\cite{jurgens2015geolocation}, evaluation after resolving the location of the unified output using a single reverse-geocoding API allowed evaluation over four geographic granularities and ensured a fair comparison using the same set of locations and avoided the mismatch of predictions based on different representations although they refer to the same location. We demonstrated how competitive geolocation models---previously proclaimed to be inferior---could compete with state-of-the-art models in terms of accuracy.

A major limitation to this work is not extending our evaluation process to network-based approaches, and more importantly recent hybrid methods that rely on deep learning. User coverage is an essential network specific metric to evaluate the percentage of test users with a predicted location~\cite{jurgens2015geolocation}. If a user does not have social ties, a network-based geolocation model will not be able to predict a location. While hybrid approaches consider network information for training, they evaluated their performance against text-based approaches using error distance measures for two reasons. First, they rely on datasets constructed by text-based research. Second, they always predict a location for a user; rely on text as a fallback if a user is disconnected. The challenge here is to address the user coverage aspect when evaluating text-based against network-based approaches. In this case, recall could be a potential metric.



\section{GeoLocEval}

Geocoding is the process of linking a document (e.g. Wikipedia article, web page, social media entity, etc.) to a location on earth. Geocoding serves a wide range of applications. With ever increasing quantities of social media content, many applications exploit such data. Examples include: dialectology (the study of geographic lexical variation of a language); regional sentiment analysis; monitoring public health; managing natural crises; and the search for eyewitnesses by journalists. Document geocoding has been an active research area over the last decade, resulting in hundreds of publications, geocoding systems and datasets~\citep{melo2017automated,zheng2018survey,mourad2018well}. Comparison of such systems share the same challenges of Twitter user geolocation. We, therefore, share our evaluation framework with the research community, hoping researchers will employ in their future geocoding research.

GeoLocEval is an open source python package~\footnote{https://bitbucket.org/amourad/geoloceval.git} to evaluate the performance of a given set of geocoding systems. The input is a list of JSON files, one for each system to be compared. Each file contains geolocations expressed in the most generic format: GPS coordinates, as in Listing~\ref{listing:input_template}. This format is compatible with the Twitter geolocation prediction shared task at the level of tweets and users~\citep{han2016twitter}, known as \textsc{Wnut}.

\begin{listing}
    \begin{minted}[autogobble, fontsize=\small]{json}
    { "doc_id": {"lon": "x", "lat": "y"}, }
    \end{minted}
    \caption{JSON Input Format}
    \label{listing:input_template}
\end{listing}

The GPS coordinates are expanded using a single geocoding API. Results are exported to a JSON file, as in Listing~\ref{listing:output_template}.

\begin{listing}
    \begin{minted}[autogobble, fontsize=\small]{json}
    {
        "483049821":
        {
            "geocoding_system_1":
            {
                "doc_id":"483049821",
                "lon":-74.0344411626724,
                "lat":40.74801738664574,
                "country":"United States",
                "county":"Hudson County",
                "state":"New Jersey",
                "city":"Hoboken",
                "error_dist":15137.622354338771
            },
        }
    }
    \end{minted}
    \caption{JSON Output Format}
    \label{listing:output_template}
\end{listing}

\subsection{Geocoding APIs}
GeoLocEval supports two of the most common geocoding APIs used in previous research:
\begin{itemize}
    \item Nominatim: a free OpenStreetMap based geocoder.
    \item GoogleV3: is a commercial API with a higher number of requests per day compared to Nominatim.
\end{itemize}
Each API supports four administrative levels, namely city, county, state and country. GeoLocEval caches all the resolved GPS coordinates to reduce the number of requests.

\subsection{Evaluation Process}
We follow the process presented in \S~\ref{sec:stndrdeval}, by first comparing systems under different evaluations and geographic granularity, next examining rank correlations of systems, and finally studying significant differences. All the generated results are exported to a text file.

\section{Conclusion and Future Work}

The work in this paper examined the effectiveness of metrics employed in the evaluation of Twitter user geolocation from three key aspects: standardized evaluation process, compensating bias due to population imbalance through micro vs macro averaging, and comprehensive statistical analysis. We proposed a practical guide to follow for an effective evaluation of each aspect based on thorough experiments and analysis encompassing fifteen geolocation models and two baselines in a controlled environment.

A recommended practical guide for any new research on Twitter user geolocation includes:
\begin{enumerate*}[label=\roman*)]
    \item creating its own dataset,
    \item sharing its geolocation model with the ability to be retrained by the research community,
    \item using a unified output format (GPS coordinates),
    \item using a single reverse-geocoding API for discrete evaluation of all the geolocation models considered,
    \item employing a combined set of evaluation metrics at the micro and macro levels,
    \item quantifying the agreement between the evaluation metrics through rank correlation and
    \item verifying the conclusions by conducting the recommended statistical significance tests.
\end{enumerate*}

This work was initially motivated by~\citet{gao2015tweet} who changed the perspective of evaluating sentiment analysis after many years of research. They argued that any study dealing with sentiment analysis is usually interested in the sentiment at the aggregate level of classes, not at the individual level. Quantification-specific evaluation metrics therefore should be used instead of classification metrics, based on the goal of the applications. Since Twitter user geolocation applications do not have a unified goal as sentiment analysis, we focused on experimental evaluation using a wide range of metrics as a vital step that leads to application-specific evaluation. For future work, we would like to investigate the evaluation of geolocation models analytically. Instead of anticipating the needs of the applications, we are interested in collaboration with domain experts, such as journalists, or humanitarians to develop the needs and evaluation metrics in the context of a specific task.

Evaluation of geolocation models on datasets with different characteristics or domains to ensure their consistent performance is a common practice. \citet{rahimi2018semi} evaluated their models on three Twitter datasets with different geographic coverage and size. \citet{mourad2017language} evaluated their model on Twitter datasets for thirteen different languages. \citet{wing2014hierarchical} evaluated their models on six datasets from different domains, namely Twitter, Wikipedia and Flickr. In this paper, we measured the statistical significance of the differences between geolocation models evaluated on the same dataset. For future work, we would like to extend our geolocation evaluation guide to include the replicability analysis for statistical significance analysis over multiple  datasets~\cite{dror2017replicability}.

\bibliographystyle{ACM-Reference-Format}
\bibliography{references}

\end{document}